\documentclass{aa} 

\usepackage{txfonts}
\usepackage{graphicx}
\usepackage{amssymb}
\usepackage{lscape}
\usepackage{natbib}
\usepackage{multirow}

\def\vsini {{v\!\sin\!i}}

\def\teff {T_{\mathrm{eff}}}

\def\feh {[Fe/H]}
\def\logg {\log g}

\def\muHz {\mu\mbox{Hz}}

\def\ds {$\delta$-Scuti}
\def\dss {$\delta$-Scuti stars}

\def\cesam {{\sc{cesam}}}
\def\graco {{\sc{graco}}}

\begin{document}

  \title{Asteroseismic analysis of the CoRoT\thanks{The CoRoT space mission, launched on December 27, 2006, has been developed and is operated by the CNES, with the contribution of Austria, Belgium, Brasil, ESA, Germany, and Spain.} $\delta$-Scuti star \object{HD\,174936}  \thanks{Table 1 is only available in electronic form
at the CDS via anonymous ftp to cdsarc.u-strasbg.fr (130.79.128.5)
or via http://cdsweb.u-strasbg.fr/cgi-bin/qcat?J/A+A/ }}
  \authorrunning{Garc\'ia Hern\'andez et al.}

  \author{A. Garc\'ia Hern\'andez\inst{1}
  \and A. Moya\inst{1}
  \and E. Michel\inst{2}
  \and R. Garrido\inst{1}
  \and J.C. Su\'arez\inst{1}
  \and E. Rodr\'iguez\inst{1}
  \and P.J. Amado\inst{1}
  \and S. Mart\'in-Ru\'iz\inst{1}
  \and A. Rolland\inst{1}
  \and E. Poretti\inst{3}
  \and R. Samadi\inst{2}
  \and A. Baglin\inst{2}
  \and M. Auvergne\inst{2}
  \and C. Catala\inst{2}
  \and L. Lefevre\inst{2}
  \and F. Baudin\inst{4}}

  \offprints{A. Garc\'ia Hern\'andez\,\email{agh@iaa.es}}

  \institute{Instituto de Astrof\'{\i}sica de Andaluc\'{\i}a (CSIC), CP3004,
Granada, Spain \and Observatoire de Paris, LESIA, CNRS, UMR 8109, Meudon,
France \and  INAF-Osservatorio Astronomico di Brera, Via E. Bianchi 46, I-23807 Merate (LC), Italy \and Institut d'Astrophysique Spatiale, UMR8617, Universit\'e Paris X, B\^at.121, 91405 Orsay, France}
  \date{Received ... / Accepted ...}

  \abstract{We present an analysis of the \ds\ star object {HD\,174936} 
            (\object{ID\,7613}) observed by CoRoT during the first short run SRc01 (27 days). A total
            number of 422 frequencies we are extracted from the light curve using standard 
            prewhitening techniques. This number of frequencies was obtained by considering a spectral significance limit of sig=10 using the software package SigSpec.
            Our analysis of the oscillation frequency spectrum 
reveals a spacing periodicity of around 52 $\mu$Hz. Although modes considered here are not in the 
asymptotic regime, a comparison with stellar models confirms that this 
signature may stem from a quasi-periodic pattern similar to the so-called large 
separation in solar-like stars.

           }

\maketitle


\section{Introduction\label{sec:intro}}


The \dss\ are intermediate-mass pulsating variables with spectral types ranging
from A2 to F0. They are located on and just off the main sequence in the lower
part of the Cepheid instability strip (luminosity classes IV \& V). 
The \dss\ are considered as particularly suitable for asteroseismic studies of
poorly known hydrodynamical processes occurring in stellar interiors, such as
convective overshoot, mixing of chemical elements, and redistribution of angular
momentum \citep{Zahn92}, to mention only the most important ones. Because of the
complexity of the oscillation spectra, their pulsating behaviour has not been
understood very well. The reader is referred to the review by \citet{Cox02} for
a more detailed description of unsolved problems in stellar pulsation physics.


The \ds\ star \object{HD\,174936} was uninterruptedly observed by CoRoT during 27 days of
observation.  
The very precise space photometry supplied by the CoRoT mission gives us the
possibility of working with a lower limit in amplitudes of $\sim$ 14 $\mu$mag. 
As a consequence of such precision, never reached by any ground-based observation campaign, we obtained a wide range of oscillation frequencies ([0.05, 100] cd$^{-1}$).
This and the huge number of detected frequencies 
for such stars will help us in 
understanding the internal structure of intermediate-mass, 
main-sequence stars.

One of the aims of the mission is that such high resolution can enable us to detect some periodicities like rotational splitting \citep{baglin2006-2}. Using the CoRoT achievements, we can carry out statistical studies and look for other possible periodicities that could help us to better
understand the oscillation spectra observed.


The paper is structured as follows: an explanation of the data analysis and the frequency content is described in Sect. \ref{sec:data}; a description of the method used for identifying the periodic spacing distribution of the extracted frequencies is given in Sect. \ref{sec:fourier}; equilibrium models and oscillation computations to check the origin of the signature noticed in the observations are described in
Sect. \ref{sec:models}; a brief discussion is carried out in Sect. \ref{sec:discussion}; and some conclusions are reported in Sect. \ref{sec:conclusions}.

\section{Data analysis \label{sec:data}}

The field \ds\ star \object{HD\,174936} was observed during the first 30-day short run
SRc01 \citep{baglin2006} of CoRoT. The final time span of the collected dataset was $\Delta$T=27.2 days, with a sampling of one point each 32 sec. This means about 73440 datapoints as the expected timeseries. However, the final dataset consisted of 66057 datapoints after removing those points considered unreliable. With this, the Rayleigh frequency resolution is (1/$\Delta$
T)=0.037 cd$^{-1}$, and an oversampling of 20 should correspond to a frequency spacing of 0.0018 cd$^{-1}$. This is equivalent to the frequency spacing in mode ``High'' in the program package Period04 \citep{lenz2005}. 

First, the data were detrended \citep{auvergne2009} performing a linear fit to the light curves. Second, Period04 was used for a preliminary inspection of the periodograms. Later, the timeseries was analysed using the computer program package SigSpec \citep{reegen2007}.

Nevertheless, the timeseries was previously investigated with Period04 for the first 20 peaks. They were identical to those achieved with SigSpec. This test has also been carried out successfully for the CoRoT target ID 123 \citep[HD\,50844,][]{poretti2009} with $\Delta$T=56.7 days and 140016 datapoints. In that case, Period04 was used to investigate the first 200 peaks, The agreement between the two methods was successful in 99\% of the cases.

\begin{figure}
\begin{center}
 \scalebox{.65}{\includegraphics{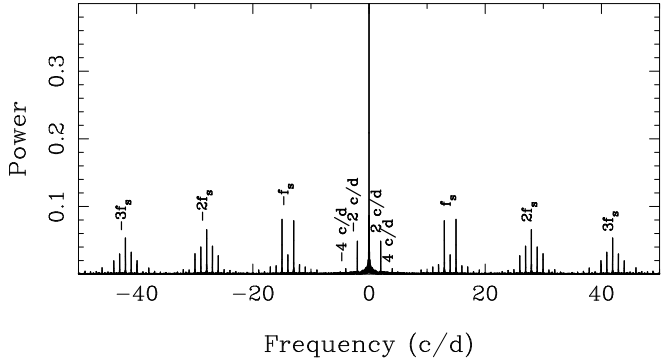}}
 \caption{Spectral window of the short-run CoRoT dataset on HD\,174936.}
 \label{fig:window}
\end{center}
\end{figure}

Figure \ref{fig:window} shows the spectral window of HD\,174936, which is typical of all the targets observed by the CoRoT satellite in the same run. As can be seen, the figure does not show the typical aliases at 1 cd$^{-1}$ or the power levels that are common for ground-based data. On the contrary, all the aliases are related to effects produced by the satellite and its orbital frequency (f$_{s}$=13.972 cd$^{-1}$), and their power levels are much lower than those usual for ground-based data. The alias peak at about 2 cd$^{-1}$ comes from the South Atlantic Anomaly (SAA) crossing, which occurred twice each sidereal day. This means the alias peaks at 2.005 and 4.011 cd$^{-1}$, respectively.

To avoid problems with power close to zero frequency, the analysis with SigSpec was carried out in the range 0.05-100 cd$^{-1}$. In the case of Period04, the limit commonly used to consider one datapoint as significant is the amplitude signal-to-noise ratio S/N$\geq$4.0. In the case of SigSpec, the parameter used for significance is ``sig'' (= spectral significance), and the limit used as default is sig=5.0. This should be equivalent to about S/N=3.8 (and sig=5.46 should be approximately equivalent to S/N=4.0) \citep{reegen2007, kallinger2008}. However, we used a much more conservative limit, and our calculations with SigSpec stopped when sig=10.0 because the corresponding S/N values, determined using Period04 on the residuals, were much lower than expected. This is probably caused by a high number of peaks still remaining among the residuals in the region of interest. This was explained in much more detail in similar recent works for other CoRoT targets \citep[HD\,50844,][HD\,49434, Rodr\'{i}guez et al., in preparation]{poretti2009}.

The limit of sig=10 was achieved after removing 422 peaks. This means a level of about 14 ppm for the smallest amplitudes. The first eleven frequencies, 
are listed in Table \ref{tab:frequencies} along with the most relevant parameters. Column 6 lists the S/N values corresponding to each peak. These S/N values were calculated using Period04 on the residual file provided by SigSpec. Each S/N value was calculated on a box of width=5 cd$^{-1}$ centred on the corresponding peak, as is usual for this type of variable \citep{rodriguez2006, rodriguez2006-2}.

\begin{table*}
  \begin{center}
   \caption{Frequencies of the highest amplitudes for \object{HD\,174936}.}
   \vspace{1em}
   \renewcommand{\arraystretch}{1.2}
   \begin{tabular}[ht!]{cccccc}
   \hline
   \hline
   Identification & Frequency ($\muHz$) & Amplitude ($mmag$) & Phase ($rad$) & Spectral significance & S/N\\
   \hline
F1         & 377.30  & 2.12  &  2.40  &    7225.844    &  659.008   \\
F2         & 412.71  & 1.02  &  2.49  &    3624.781    &  336.581   \\
F3         & 414.62  & 0.72  & -0.41  &    2155.327    &  236.220   \\
F4         & 360.08  & 0.57  &  2.67  &    1784.588    &  174.791   \\
F5         & 339.22  & 0.55  &  1.18  &    1714.847    &  169.363   \\
F6         & 367.96  & 0.53  & -0.82  &    1862.847    &  164.484   \\
F7         & 321.92  & 0.36  & -1.36  &     908.218    &   111.772  \\
F8         & 387.63  & 0.32  & -2.55  &     825.756    &   103.085  \\
F9         & 359.47  & 0.30  &  2.23  &     713.637    &    92.766  \\
F10       & 385.89  & 0.29  & -0.05  &     725.392     &   93.201    \\
F11       & 158.01  & 0.28  & -2.44  &     707.305     &   85.501    \\
     \hline
   \hline
 \end{tabular}
   \label{tab:frequencies}
  \end{center}
\end{table*}
 
\begin{figure}
\begin{center}
 \scalebox{.45}{\includegraphics{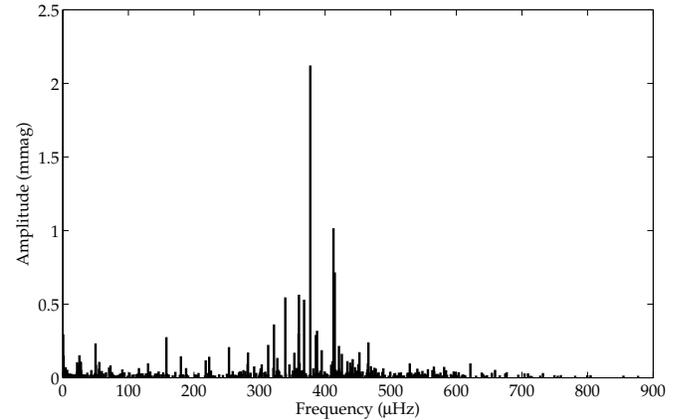}}
 \caption{The 422 frequencies exctracted for HD\,174936.}
 \label{fig:spectrum}
\end{center}
\end{figure}

The range of statistically significant detected frequencies
goes from some value close to zero up to about 70 cd$^{-1}$, i.e. $800\,\muHz$ (1 cd$^{-1}$ = 11.57$\muHz$).
But those with the highest amplitudes are grouped around 35 cd$^{-1}$ ($400\,\muHz$, see Fig. \ref{fig:spectrum}). 
In the following theorical analysis, we work in $\muHz$.

\section{Looking for periodic patterns in the extracted frequency set
\label{sec:fourier}}

Mode identification, mainly for \dss, is the present bottleneck for the
progress in asteroseismology. 
Existing mode identification techniques are based on spectroscopy or
multi-colour photometry not available in CoRoT data. Considering that we have 
unprecedented, rich spectra, we seek to reveal characteristic 
structures in the frequencies 
distribution and to start with quasi-regular spacings.

We consider the frequency set as a series of 
Dirac's $\delta$'s of equal amplitude. Then, the technique is simply to use the 
list of frequencies with unit amplitude and to normalise its Fourier transform 
to unity at zero frequency spacing.  
If we have a given
periodicity in the frequency set ($\delta f$), then its Fourier transform will show a
clear periodic structure on the corresponding inverse scale. 

\begin{figure*}
\begin{center}
 \scalebox{.70}{\includegraphics{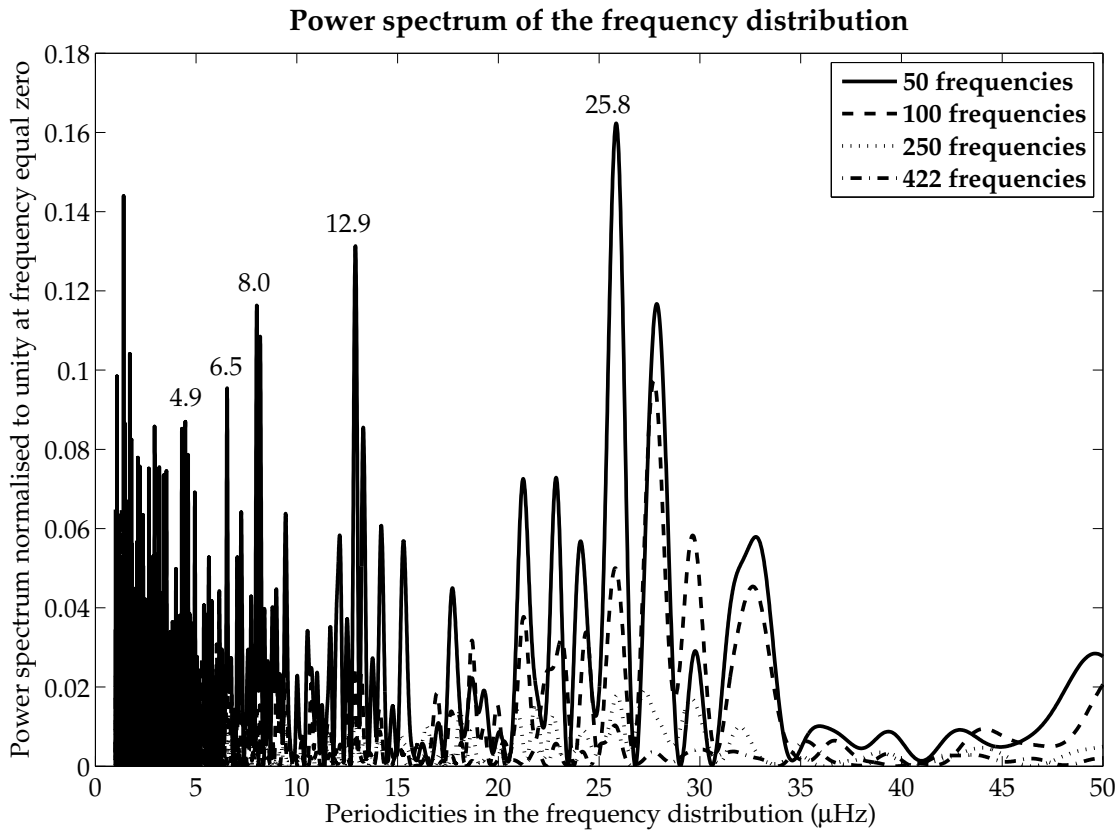}}
 \caption{Power spectrum for various subsets of frequencies, selected by
          amplitude. The solid line represents the power spectrum when the highest 50
          frequencies are selected; the dashed line corresponds to the highest 100 frequencies;
          the dotted line, for 200; and dot-dashed line for all of them. The peak
          corresponding to the large separation (25.8 $\mu\mbox{Hz}$) and its
          sub-multiples are labelled (see text).}
 \label{fig:large-sep}
\end{center}
\end{figure*}

We started this analysis assuming the hypothesis that, selecting a subset with the highest
peaks of the entire frequency set, we are actually considering mainly the modes of the lowest
$\ell$ values. The visibility of the modes
decreases approximately as $\ell^{-2.5}$ or $\ell^{-3.5}$, depending whether
the $\ell$ degree is odd or even \citep{dziembowski1977}. 
We analysed several subsets with an
arbitrary number of frequencies but always including the highest peaks. In Fig.~\ref{fig:large-sep}
the results are given for different sets of peaks. 
It can be seen that the 
more frequencies in the subset, the flatter the peaks, although 
the local maxima remain at the same position. This indicate that, when we only select the highest peaks, we are enhancing some periodicities corresponding to a few $\ell$ values. While, adding frequencies in the analysis, periodicities of all modes are powered and no pattern can be significantly distinguished from others.

When we select 50 frequencies, a periodic pattern 
can be recognised. Furthermore, the clear signature of a Dirac's comb is also present at 
25.8, 13, 8, 6.5, and 4.9 $\muHz$. That
indicates that we probably have a Dirac's comb of 25.8 $\muHz$, which is the
lowest divisible number. The appearance of these peaks can be a clue to the confirmation that
they are being caused by a periodic structure in the frequency set. Other peaks around 
25.8 $\muHz$ can be explained by the separation not being strictly periodic but quasi-periodic, as can be expected for these modes (see Sect. \ref{sec:models}). This can be the imprint of a large separation-like structure around 52 $\muHz$.\footnote{Frequency values for $\ell$ = 1 modes are placed around the centre of two consecutive $\ell$ = 0 modes, so we expect to find a minimun periodicity of the half value of the large separation.}

On the other hand, the rotational velocity measured for this star is high ($v\sin i \sim 170$ km/s, 
see Sect. \ref{sec:models}) giving a minimum rotational splitting of $\sim 20\ \mu$Hz. 
It is then possible that the observed frequency pattern does not come from the large separation but from the rotational splitting. The pattern due to rotational effects should be observed only with $\ell\geq1$ and a statistically significant number of $m\neq$ 0 values. 
However, because we 
mainly select the lowest $\ell$ values, the large separation is more probable than the rotational splitting. 
In the next section, we discuss the validity of this hypothesis using
a theoretical model representative of the star.

The other peaks in Fig.~\ref{fig:large-sep} probably come from periodic structures produced by the rotation and/or
the small separation, mainly occupying the lowest part of the power spectrum. 
Works are in progress for understanding the meaning of these peaks.

\section{Periodic structures in theoretical spectra \label{sec:models}}

The periodic pattern structures are familiar in solar-like oscillators, which have reached
the asymptotic regime, but not in \dss, which pulsate around the fundamental
radial mode. 
To understand the physical origin of the present observed periodic pattern shown
in the previous section, we computed a representative model of this star.

\subsection{Physical parameters \label{subsec:parameters}}

We consider the physical parameters of the star \object{HD\,174936}, which 
are listed in Table \ref{tab:parfis}. These values are taken from ``CorotSky Database''
\citep{Charpinet2006}. $\teff$, $\logg$, and $\feh$ were derived from
Str\"omgren photometry \citep{Hauck1998},  while the rotational velocity ($v\sin
i = 169.7 km/s$) was determined from high-resolution spectroscopy. The
spectrum was taken in June 2, 2004 using the FEROS spectrograph attached to the
ESO 2.2 m-telescope at La Silla (Chile). It was obtained in the framework of the
mission preparation and is available at the GAUDI archive \citep{solano2005}.
On the other hand, it has been shown that rapid rotation should be considered
when calculating the stellar physical parameters from photometry. \cite{miher99}
propose a method for determining the effects of rotation and geometry (angle of
inclination) on photometric parameters for stars in clusters. For $\delta$-Scuti stars, that method
was further developed by \cite{Pe99}, showing that uncertainties of around
100-150 K in effective temperature, and $\sim$ 0.10 dex in $logg$, can be found
for moderately rotating stars. That result was later confirmed by \cite{sua02aa}.
In the present case, considering the absence of additional information on the
inclination angle of the star, uncertainty boxes of 200 K in the HR diagram in
$T_{eff}$ and $\sim$ 0.2 dex in $logg$ were adopted. An uncertainty of 
$\sim$ 0.2 in metallicity was taken.

This is a main-sequence intermediate-mass star ($\sim$1.6 M$_{\odot}$). The hydrogen abundance in the core is X$_{c}\sim0.3$. But it is younger than the star studied in \citet{poretti2009}.
\subsection{Theoretical models \label{subsec:the_models}}

Using the physical parameters measured for the star and given in the previous subsection as input, we computed a theoretical
model representative of the star. The evolutionary code {\cesam} \citep{Morel97,morel2008} and the pulsation code {\graco}
\citep{Moya04,graco08} were
used as numerical codes to calculate frequencies, growth rates and other
physical quantities.

\begin{figure}
\begin{center}
 \scalebox{.30}{\includegraphics{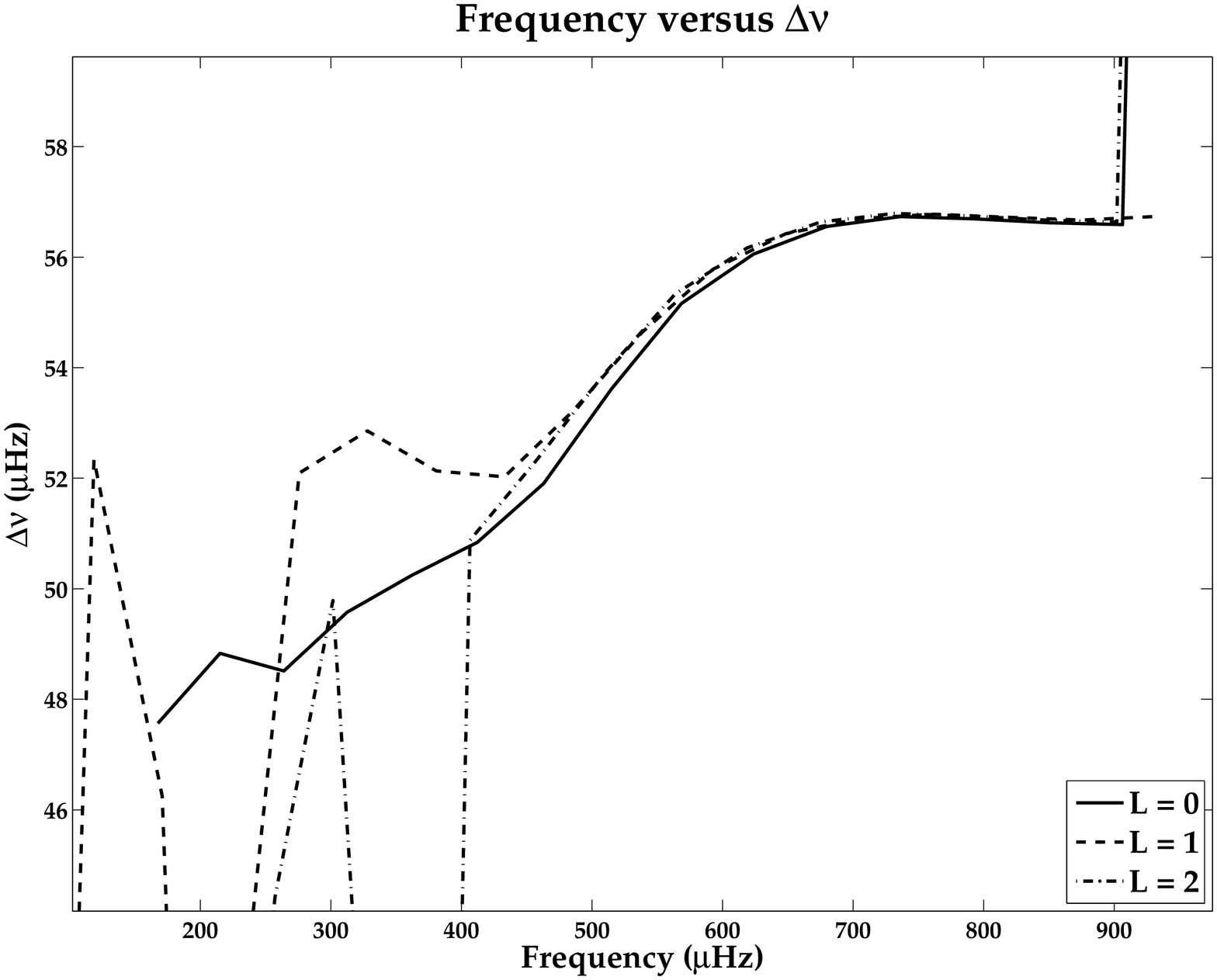}}
 \caption{Frequency dependance of the large separation for the
          three lowest values of $\ell$: 0, 1, and 2. Solid line is for $\ell=$0, dashed
         line for $\ell=$1, and dot-dashed line for $\ell=$2. 
         Large separation is then estimated using a Fourier transform method (see
         Sect.~\ref{sec:fourier}).}
 \label{fig:mean-large-sep}
\end{center}
\end{figure}

For the model, we computed the
oscillation spectrum (radial and non-radial), from $\ell=0$ to 3, in the
range of observed frequencies.
Then, we calculated the large separation defined as
$\Delta\nu=\nu_{n+1,\ell}-\nu_{n,\ell}$. In Fig.~\ref{fig:mean-large-sep} the large
separation calculated this way is depicted. It is quite evident that the large
separation does not vary much but is bounded within a frequency range of 
10 $\muHz$. In this particular case, the two discrepant points belonging
to $\ell=2$ correspond to an avoided crossing. Similar results are
expected for other higher $\ell$ values.

\begin{figure}
\begin{center}
 \scalebox{.40}{\includegraphics{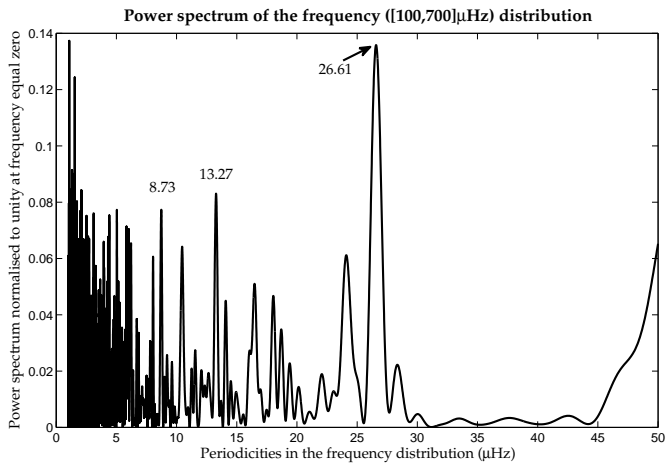}}
 \caption{Power spectrum of the oscillation frequencies of a model obtained using the technique 
         described in Sect.~\ref{sec:fourier}. We use only the range of observed data set, i.e., up to 700 $\muHz$. A clear peak (and sub-multiples) 
          is observed at 26.61 $\muHz$, which corresponds to a large separation of $\sim$53 $\muHz$.}
 \label{fig:large-sep-theoric}
\end{center}
\end{figure}

However, this is not critical
for the detection of regularities using Fourier transform due to its
statistical character.
In Fig.~\ref{fig:large-sep-theoric} we show an identical analysis to the one for
the observed frequencies (see previous section). It can be seen 
that we have similar results for the theoretical frequencies
for a given representative model (even in a nonrotating model). This indicates that such regularities, 
even in the non-asymptotic regime, may be discerned by the Fourier transform
technique explained in Sect.~\ref{sec:fourier}. This opens a new window
in the analysis of \dss.

\begin{table}
  \begin{center}
   \caption{Physical parameters of the star with its uncertainties to
construct the theoretical model \citep{Charpinet2006}.}
   \vspace{1em}
   \renewcommand{\arraystretch}{1.2}
   \begin{tabular}[ht!]{ccccc}
   \hline
   \hline
   Star ID & $\teff$ (K) & $\logg$ & $\feh$ & $\vsini$ (km/s) \\
     \hline
     \object{HD\,174936} & 8000$\pm$200 & 4.08$\pm$0.2 & -0.32$\pm$0.2 & 169.7
\\
     \hline
     \hline
   \end{tabular}
   \label{tab:parfis}
 \end{center}
\end{table}

\begin{figure}
\begin{center}
 \scalebox{.40}{\includegraphics{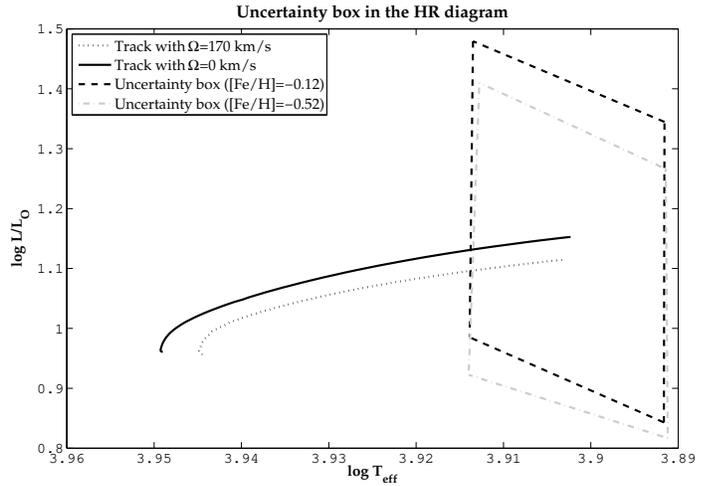}}
 \caption{HR diagram containing two uncertainty boxes corresponding to the $1\sigma$ extremes of the metallicity ([Fe/H] = -0.12 and -0.52), with two evolutionary tracks. They correspond to the central model (M = 1.63 M$_{\odot}$) at the central metallicity ([Fe/H] = -0.32), one with rotation and the other without.}
 \label{fig:uncertainty}
\end{center}
\end{figure}

\section{Discussion \label{sec:discussion}}

We have shown that a periodic pattern, possibly corresponding to large separation, is found in the frequency spectrum observed for the star (see Sect. \ref{sec:fourier}). We have corroborated it through a theoretical model (see previous section). Clearly, the large separation found in the studied regime does not have the same physical meaning as found in the asymptotic regime \citep{JCD03}. However, it is dependent on the internal structure of the star, and we can study how this signature is robust and sensitive to evolution or mass, in order to open up new perspectives about its diagnostic power.

To do so, we constructed seismic
models that lie within the uncertainties of the physical parameters measured for this star. That is, we computed models to reproduce the HR locations corresponding to the four corners ($1\ \sigma$) and the centre
of the uncertainty box. After that, for each equilibrium model, we computed its 
oscillations. This procedure is followed for models built with the observed 
metallicity and the two maximum errors ($1\ \sigma$) considered. Then, we reduced the corners of the uncertainty box to half and computed the corresponding models. Finally, we calculated the large separation for all them. The uncertainty boxes ($1\ \sigma$) for the $1\ \sigma$ extremes of the metallicity are depicted in Fig. \ref{fig:uncertainty}. Two evolutionary tracks of the central model (M = 1.63 M$_{\odot}$) with and without rotation are plotted, too.

We can see in Table \ref{tab:models-instab} the results of all these calculations. The large separation is close to the value found in the observed frequency set for the central models at each metallicity. Typically, it is higher for less massive models, and when reducing the half-bottom area of the uncertainty box it varies faster than the half-upper one.

\begin{table*}
  \begin{center}
   \caption{Models of the uncertainty boxes for each metallicity calculated with mass and large separation.}
   \vspace{1em}
   \renewcommand{\arraystretch}{1.2}
   \begin{tabular}[ht!]{ccccccc}
   \hline
   \hline
\multicolumn{3}{c}{First box}& &\multicolumn{3}{c}{Second box} \\
$\feh$ & $M$ ($M_{\odot}$) & $\Delta\nu$ ($\mu Hz$) & & $\feh$ &
$M$ ($M_{\odot}$) & $\Delta\nu$ ($\mu Hz$)  \\
     \hline
      \multirow{5}{*}{-0.32} & 1.63 & 53 &
&\multirow{5}{*}{-0.32} & 1.63 & 53  \\
      & 1.89 & 37  & & & 1.77 & 44  \\
      & 1.77 & 38  & & & 1.71 & 44  \\
      & 1.51 & 82  & & & 1.58 & 60   \\
      & 1.43 & 71  & & & 1.54 & 58  \\
     \hline
     \multirow{5}{*}{-0.52} & 1.51 & 54  &
&\multirow{5}{*}{-0.52} & 1.51 & 54  \\
      & 1.76 & 39  & & & 1.64 & 44  \\
      & 1.62 & 41  & & & 1.58 & 45  \\
      &  1.42 & 74  & & & 1.48 & 60  \\
      &  1.33 & 73  & & & 1.43 & 60  \\
     \hline
      \multirow{5}{*}{-0.12} & 1.76 & 51  &
&\multirow{5}{*}{-0.12} & 1.76 & 51  \\
      & 2.05 & 37  & & & 1.92 & 43  \\
      & 1.90 & 39  & & & 1.85 & 43  \\
      & 1.66 & 71  & & & 1.73 & 58  \\
      & 1.55 & 73  & & & 1.67 & 58  \\
     \hline
   \hline
 \end{tabular}
   \label{tab:models-instab}
  \end{center}
\end{table*}

\begin{figure}
\begin{center}
 \scalebox{.39}{\includegraphics{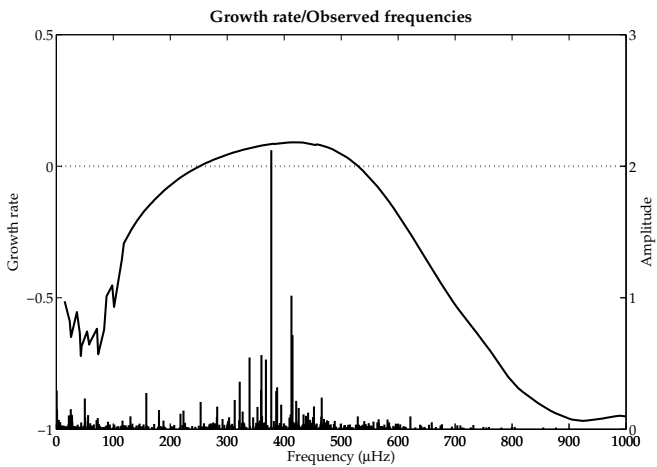}}
 \caption{Frequency versus growth rate and amplitude for the 1.63 M$_{\odot}$ model. Growth rate (solid line,
left axis) shows the stability range ($<$ 0), and one vertical line of the length
of the amplitude for each frequency is represented (right axis).}
 \label{fig:growth}
\end{center}
\end{figure}

On the other hand, we computed the instability range for these models (see Fig. \ref{fig:growth}). Some models covers the range (around 400 $\muHz$) of the highest amplitudes in the frequency set, others do not. But not one of them gives the necessary instability range to represent the full range in the observations. It is assumed that observed frequencies ([0.5, 900] $\muHz$), at least in ground-based data, must be within the instability range. However, as far as we know, no theory allows such a huge range of excited frequencies, even including the time-dependent convection theory. This is a new challenge for theory that should be improved in order to predict the instability range observed in the new era of space telescopes.

As seen, the observed frequency set contains a lot of information (small separation, rotational splitting, etc.), and work is progressing. We hope that such a large number of frequencies will allow us to obtain precise models representative of this star and to improve our understanding of the interiors of intermediate-mass stars.

\section{Summary \& conclusions \label{sec:conclusions}}

This work has presented an analysis of the \ds\ star object {HD\,174936} 
(\object{ID\,7613}), observed by CoRoT during the first short run SRc01 (27 days). A total
number of 422 frequencies (which reach up $700\,\muHz$) are extracted from the light curve using 
standard prewhitening techniques. This represents one of the largest number of detected 
frequencies for a \ds\ ever obtained so far.

We combined the classical seismic analysis with the use of statistical
properties of the modes distribution in the observed frequency range. In particular, we have 
found periodic patterns in the observed frequency spectrum, which were not expected
for this kind of pulsating star.
We find a frequency distribution of the modes that seems to
correspond to the large separation, which is about 52 $\muHz$. This result is supported
by an equivalent analysis performed with a representative model of the star.

All these results provide new prospects for the asteroseismology of \dss.
In particular, the large separation found for this star opens the possibility of performing
similar studies to those for solar-like stars. Although we have not explored the
low-frequency region of Fig.~\ref{fig:large-sep}, we cannot discard the possibility that relevant
information concerning rotation and small differences are hidden in this peak forest. 

\acknowledgements {A.G.H., A.M., J.C.S., R.G., E.R., P.J.A., S.M.R. and A.R. acknowledge
support from the Spanish ``Plan Nacional del Espacio" under project
ESP2007-65480-C02-01. A.G.H. acknowledges support from a ``FPI" contract of the
Spanish Ministry of Science and Innovation. J.C.S. acknowledges support from the
``Consejo Superior de Investigaciones Cient\'{\i}ficas" by an ``I3P" contract financed by the
European Social Fund. P.J.A. acknowledges financial support from a ``Ram\'on y
Cajal" contract of the Spanish Ministry of Science and Innovation. A.M. 
acknowledges financial support from a ``Juan de la Cierva" contract of the 
Spanish Ministry of Science and Innovation. EP acknowledges  support from the Italian
ESS project, contract ASI/INAF I/015/07/0, WP 03170.}

\bibliography{11932.bib}
\bibliographystyle{aa}

\end{document}